\documentclass[12pt]{article}
\usepackage[pdftex]{graphicx}

\begin{document}

\title{
Numerical Simulation of the SVS 13\\
Micro-Jet and Bow Shock Bubble
}
\author{
{\em Short title: Simulation of SVS 13 Micro-Jet and Bubble}\\ \ \\
Carl~L. Gardner$^1$,
Jeremiah~R. Jones$^1$,
and Klaus~W. Hodapp$^2$\\
$^1$School of Mathematical \& Statistical Sciences\\
Arizona State University, Tempe AZ 85287\\
carl.gardner@asu.edu, jrjones8@asu.edu\\
$^2$Institute for Astronomy\\
University of Hawaii, Hilo HI 96720\\
hodapp@ifa.hawaii.edu
{\em  numerical --- stars: jets} }

\date{}

\maketitle
\thispagestyle{empty}

\begin{abstract}
Numerical simulations are performed using the WENO method of the SVS
13 micro-jet and bow shock bubble which reproduce the main features
and dynamics of Keck Telescope/OSIRIS velocity-resolved integral field
spectrograph data: an expanding cooler bow shock bubble, with the
bubble center moving at approximately 50 km~s$^{-1}$ with a radial
expansion velocity of 11 km~s$^{-1}$, surrounding the fast hotter jet,
which is propagating at 156 km~s$^{-1}$.  Contact and bow shock waves
are visible in the simulations from both the initial short jet pulse
which creates the nearly spherical bow shock bubble and from the fast
micro-jet, while a terminal Mach disk shock is visible near the tip of
the continuous micro-jet, which reduces the jet gas velocity down to
the flow velocity of the contact discontinuity at the leading edge of
the jet.  At 21.1 yr after the launch of the initial bubble pulse, the
jet has caught up with and penetrated almost all the way across the
bow shock bubble of the slower initial pulse.  At times later than
about 22 yr, the jet has penetrated through the bubble and thereafter
begins to subsume its spherical form.  Emission maps from the
simulations of the jet---traced by the emission of the shock excited
1.644 $\mu$m [Fe~II] line---and bow shock bubble---traced in the lower
excitation 2.122 $\mu$m H$_2$ 1--0 S(1) line---projected onto the
plane of the sky are presented, and are in good agreement with the
Keck observations.
\end{abstract}

\section{Introduction}

Keck Telescope laser adaptive-optics integral field spectroscopy with
OSIRIS by Hodapp \& Chini\footnote{The current paper is a companion
  paper to this work, where the observational history of SVS 13 is
  summarized in detail and more Keck Telescope images, covering the
  period 2011 August 21--2013 November 23, are presented and
  discussed.}  (2014) of the innermost regions of the NGC 1333 SVS 13
outflow (which forms the system of Herbig-Haro objects~7--11) revealed
a bright micro-jet traced by the emission of shock excited [Fe~II].
In addition, a series of bow shock bubbles and fragments of bubbles
beyond this micro-jet were observed in the lower excitation H$_2$ 1--0
S(1) line.

The kinematic age of the youngest bubble is slightly older than the
last observed photometric outburst of SVS 13 in 1990, and consistent
with that event launching the bubble with some subsequent deceleration
of its expansion.

Jet outflows emanating from young stars typically are comprised of a
series of individual shock fronts that arise from changes in the jet
velocity as a result of repetitive eruptive events (see the review by
Reipurth \& Bally (2001) and references therein).  This general
framework can be applied to SVS 13: the FUor-like photometric outburst
in 1990 produced a short-lived pulse of jet activity which formed a
bow shock bubble as it ran into slower moving material ejected prior
to the outburst event.  Subsequent to the formation of the bubble, the
SVS 13 stellar source ejected a fast continuous micro-jet that caught
up with and pierced the bubble, partially incorporating and destroying
the bubble.  In SVS 13, this process is currently repeating itself
about every 30 yr, creating the series of bubble fragments that make
up the string of Herbig-Haro objects.

The creation of a series of bubbles and the changes in outflow
direction indicate that the accretion disk of SVS 13 is precessing.
The 30 yr cycle of bubble pulses may correspond with the orbit of an
as yet unobserved close binary companion, and/or to revolving
inhomogeneities in the accretion disk.

In this investigation, we present the first numerical simulations of
the SVS 13 micro-jet and most recent bow shock bubble, using the WENO
method. These simulations reproduce the main features and dynamics of
the Keck data: an expanding cooler bow shock bubble, with the bubble
center moving at approximately 50 km~s$^{-1}$ with a radial expansion
velocity of 11 km~s$^{-1}$, surrounding the fast hotter jet, which is
propagating at 156 km~s$^{-1}$.  Emission maps from the simulations of
the jet---traced by the emission of the shock excited 1.644 $\mu$m
[Fe~II] line---and bow shock bubble---traced in the lower excitation
2.122 $\mu$m H$_2$ 1--0 S(1) line--- projected onto the plane of the
sky are also presented, and are in good agreement with the Keck
observations.

The only other known examples of a young stellar object outflow with
pronounced bubble structures are XZ Tauri (Krist et al.\ 2008) and
IRAS 16293--2422 (Loinard et al.\ 2013).  The first numerical
simulations of bow shock bubbles originated and penetrated by jet
pulses (in XZ Tauri) were presented in Krist et al.\ (2008) and
Gardner \& Dwyer (2009), using our WENO code.  These simulations
modeled a faster, more rapidly pulsing proto-jet which produced more
rapidly expanding and overlapping bubbles.

\section{Numerical Method and Radiative Cooling}

We applied the WENO method---a modern high-order upwind method---to
simulate the expanding bow shock bubble and jet in SVS 13, including
the effects of atomic and molecular radiative cooling.

We use a third-order WENO (weighted essentially non-oscillatory)
method (Shu 1999) for the simulations (Ha et al.\ 2005).  ENO and WENO
schemes are high-order finite difference schemes designed for
nonlinear hyperbolic conservation laws with piecewise smooth solutions
containing sharp discontinuities like shock waves and contacts.
Locally smooth stencils are chosen via a nonlinear adaptive algorithm
to avoid crossing discontinuities whenever possible in the
interpolation procedure.  The weighted ENO (WENO) schemes use a convex
combination of all candidate stencils, rather than just one as in the
original ENO method.

Our WENO code is parallelized using OpenMP and MPI, and can simulate
both cylindrically symmetric flows (Figures~2--11 below) as well as
fully three-dimensional flows (Figures~12 and~13).

The equations of gas dynamics with radiative cooling take the form of
hyperbolic conservation laws for mass, momentum, and energy:
\begin{equation}
	\frac{\partial \rho}{\partial t} + 
	\frac{\partial}{\partial x_i} (\rho u_i) = 0
\label{n}
\end{equation}
\begin{equation}
	\frac{\partial}{\partial t} (\rho u_j) + \frac{\partial}{\partial x_i}
	(\rho u_i u_j) + \frac{\partial P}{\partial x_j}
	= 0
\label{p}
\end{equation}
\begin{equation}
	\frac{\partial E}{\partial t} + \frac{\partial}{\partial x_i}
	(u_i (E + P)) = - C(n, T)
\label{E}
\end{equation}
where $\rho = m n$ is the density of the gas, $m$ is the mass of the
gas atoms or molecules (largely either $m_H$ or $2 m_H$ where $m_H$ is
the mass of H), $n$ is the number density, $u_i$ is the velocity,
$\rho u_i$ is the momentum density, $P = n k_B T$ is the pressure,
$k_B$ is Boltzmann's constant, $T$ is the temperature, and
\begin{equation}
	E = \frac{3}{2} n k_B T + \frac{1}{2} \rho u^2
\end{equation}
is the energy density.  Indices $i,~ j$ equal 1, 2, 3, and repeated
indices are summed over.

Radiative cooling of the gas is incorporated through the right-hand
side $- C(n, T)$ of the energy conservation equation~(\ref{E}), where
\begin{equation}
        C(n, T) = \left\{ \begin{array}{ll}
        n^2 \Lambda(T) & T \ge 8000~{\rm K,~ atomic~cooling~only}\\
        n W(n, T) & 100~{\rm K} \le T < 8000~{\rm K,~ H}_2~{\rm cooling~only}\\
	\end{array} \right.
\end{equation}
with the model for $\Lambda(T)$ taken from Figure~8 of Schmutzler \&
Tscharnuter (1993) for atomic cooling (encompassing the relevant
emission lines of the ten most abundant elements H, He, C, N, O, Ne,
Mg, Si, S, and Fe in the interstellar medium (ISM), as well as
relevant continuum processes) and the model for $W(n, T)$ from
Figure~4 of Le Bourlot et al.\ (1999) for H$_2$ molecular cooling.  In
these two references, the cooling functions are calculated in a
thermally self-consistent way---{\em not}\/ assuming thermal
equilibrium---involving detailed energy balancing and time-dependent
stages of ionization.  $\Lambda(T)$ incorporates sub-dominant heating
in addition to dominant cooling.  (Both atomic and molecular cooling
are actually operative between 8000~K $\le T \le$ 10,000~K, but atomic
cooling is dominant in this range.)  Note that $C(n, T)$ is
discontinuous at $T$ = 8000~K; however this does not cause any
mathematical or numerical problems, since mathematically the
homogeneous gas dynamical equations already support stronger
discontinuities (discontinuous shock and contact waves), and since
numerically $-C(n, T)$ is integrated over a timestep ({\em not}\/
differentiated in time or space).

We assume that the ambient gas (including the cool bow shocked ambient
of the bubble) is almost entirely H$_2$ throughout the simulation, and
use part~(a) of Figure~4 of Le Bourlot et al.\ (1999) with
$n(H)/n(H_2) = 0.01$.  The bubble bow shock running into the
stationary ambient molecular hydrogen at 11 km s$^{-1}$ is capable of
exciting the 1--0 S(1) line emission while not dissociating the H$_2$
(see for example Draine 1980).  We assume that the jet gas and the
immediately adjacent, hot, turbulently mixed, bow shocked jet/ambient
gas (above 8000~K) are predominantly H, with the standard admixture of
the other nine most abundant elements in the ISM.

Below we calculate emission maps from the 2D cylindrically symmetric
simulations of the SVS 13 jet and bow shock bubble projected onto the
plane of the sky, at a distance of $R$ = 235 pc with a declination
angle of 30$^\circ$ with respect to the line of sight.  The jet is
best traced by the emission of the shock excited 1.644 $\mu$m [Fe~II]
line, while the bow shock bubble is best traced in the lower
excitation 2.122 $\mu$m H$_2$ 1--0 S(1) line (Hodapp \& Chini 2014).

To calculate the 1.644 $\mu$m [Fe~II] and 2.122 $\mu$m H$_2$ emission,
we post-processed the computed solutions using emissivities extracted
from the astrophysical spectral synthesis package {\em Cloudy}\/
(version 13.03, last described by Ferland et al.\ 2013).  For the
[Fe~II] emission line, we calculated the emissivity $\epsilon_{1.644
  \mu m}$ as a function of $n$ and $T$.  Similarly for the H$_2$
emission line, we calculated $\epsilon_{2.122 \mu m}$ as a function of
$n_{H_2}$ (assuming that the ambient gas is almost entirely H$_2$
below 8000~K) and $T$.  In practice, we calculated a table of values
for $\log_{10}(\epsilon_{line})$ on a grid of $(\log_{10}(n),
\log_{10}(T))$ values relevant in the simulations to each line, and
then used bilinear interpolation in $\log_{10}(n)$ and $\log_{10}(T)$
to compute $\log_{10}(\epsilon_{line})$.

Surface brightness $S_{line}$ in any emission line is then calculated
by integrating along the line of sight through the jet and bubble
\begin{equation}
        S_{line} = \frac{\int \epsilon_{line}(n, T) dl}{4 \pi R^2}
\end{equation}
and converting to erg cm$^{-2}$ arcsec$^{-2}$ s$^{-1}$, with $R$ = 235
pc and 20 mas = 4.7 AU for SVS 13.  For details and validation of the
emission line calculations, see C.L.G., J.R.J., \& P.~B. Vargas (in
preparation).

An alternative approach to ionization effects in radiative cooling of
astrophysical jets is outlined in Tesileanu et al.\ (2014).

\section{Numerical Simulations}

Parallelized simulations of the SVS 13 bow shock bubble and jet were
performed on a $1200 \Delta z \times 400 \Delta r$ grid spanning $3.3
\times 10^{10}$ km by $2.2 \times 10^{10}$ km (using the cylindrical
symmetry) for the cylindrically symmetric simulations in Figures~2--11
and a $500 \Delta x \times 500 \Delta y \times 750 \Delta z$ grid
spanning $2 \times 10^{10}$ km by $2 \times 10^{10}$ km by $3 \times
10^{10}$ km for the fully 3D simulations in Figures~12 and~13.  The
initial bubble pulse and the jet are emitted at the center $(r,z) =
(0,0)$ or $(x,y,z) = (0,0,0)$ of the left boundary of the
computational grid in Figures~2--13, with a jet diameter of $5 \times
10^8$ km, propagating to the right along the $z$ axis.  All the
simulation images in Figures~2--13 are more or less zoomed in, to best
display the results.

The Keck Telescope image in Figure~1 shows the jet propagating down
and to the left towards the observer at an angle of about
20--30$^\circ$ with respect to the line of sight.  The Keck data show
an expanding cooler bow shock bubble, with the bubble center moving at
approximately 50 km~s$^{-1}$ with a radial expansion velocity of 11
km~s$^{-1}$, surrounding the fast hotter jet, which is propagating at
156 km~s$^{-1}$.  In the Keck image, the bubble has a diameter of
approximately $10^{10}$ km (reproduced in the simulations).
Kinematics then suggests that the bubble is 21 yr old in Figure~1
(imaged in 2011), with its emission corresponding to a photometric
outburst of the SVS 13 stellar source in 1990.

\begin{table}[htbp]
\caption{Parameters for the jet, bow shock bubble, and ambient gas.}
\begin{centering}
\begin{tabular}{lll} \hline \hline 
{\em jet} & {\em bubble pulse} & {\em ambient gas} \\ \hline 
$\rho_j$ = 1000 H cm$^{-3}$ & $\rho_b$ = 1000 H cm$^{-3}$ & 
  $\rho_a$ = 1000 H$_2$ cm$^{-3}$ \\ 
$u_j$ = 156 km s$^{-1}$ & $u_b \approx 50$ km s$^{-1}$ & $u_a$ = 0 \\ 
$T_j$ = 1000 K & $T_b$ =  1000 K & $T_a$ = 500 K \\ 
$c_j$ = 3.7 km s$^{-1}$ & $c_b$ = 3.7 km s$^{-1}$ & $c_a$ = 2.6 km s$^{-1}$ \\ 
\hline
\end{tabular}
\label{sim-params}
\end{centering}
\end{table}

The initial pulse and jet are presumably denser and cooler than the
far-field ambient gas, but we believe the pulse and jet are
propagating into previous outflows from the star, so they may in fact
be less dense and hotter than the near-field ambient wind from the
star.  For the simulations, we took the pulse and jet to be half as
dense as the near-field ambient, and twice as hot.  If the pulse that
generates the bubble is denser than the ambient, extremely high
temperatures are generated ($\gg$ 100,000~K), which disagree with the
Keck data.  By assuming the bubble and jet are less dense than the
immediate ambient gas, temperatures are kept well below 10,000~K for
the most part.

In the numerical simulations, the initial thin cylindrical-disk jet
pulse of supersonic gas, which creates the nearly spherical bow shock
bubble, has a density of 1000 H cm$^{-3}$ and a temperature of 1000~K
(see Table~\ref{sim-params}).  In the simulations, the center of the
bow shock bubble propagates at an average velocity of approximately 50
km~s$^{-1}$ over the first 23 yr.  A short initial pulse of duration
0.08 yr was allowed to propagate for 18.5 yr to form the bow shock
bubble (see Figure~2), and then the fast jet was turned on and allowed
to propagate to 20.4, 21.1, 22.3, 26.1, and 30 yr respectively in
Figures~3--11.  (Figures~2--11 show temperature and density cross
sections of the simulations in a plane containing the $z$ axis, and
surface brightness of the jet and bubble projected onto the plane of
the sky.)  The jet inflow is propagating at 156 km~s$^{-1}$, with the
same density and temperature as the initial pulse.  The jet is
surrounded by a strong bow shock plus bow shocked envelope in
Figures~3--11.  The undisturbed ambient density is 1000 H$_2$
cm$^{-3}$, with a temperature of 500~K, so that the different gases
are initially pressure matched.  The jet is propagating at Mach 60
with respect to the soundspeed in the heavier ambient gas and Mach 42
with respect to the soundspeed in the jet gas.

The Keck image in Figure~1 shows an asymmetric bow shock bubble, with
a much weaker shock on one side (also see Figures~1 and~4 in Hodapp \&
Chini (2014)).  This weakening could be caused by the jet running into
heavier and cooler gas from a previous outflow on that side, as in
Figure~12.  If the previous outflow on that side were lighter and
hotter, the bow shock bubble would expand and become hotter, as in
Figure~13.

\section{Discussion}

The surface of the bow shock bubble remains $\sim$ 1000~K throughout
the simulations, while some parts of the narrower bow shocked envelope
of the jet remain around 10,000~K in Figures~3--11.  The temperature
of the terminal Mach disk of the jet is around 100,000~K throughout.

Contact and bow shock waves are visible in the simulations in
Figures~2--11 from both the initial short jet pulse which creates the
nearly spherical bow shock bubble and from the fast jet.  The initial
pulse which creates the nearly spherical bow shock bubble (see
Figure~2) is visible as a contact wave in the figures---as a spherical
blob in density in Figures~2--4, and as a conical ``wing'' as the
initial short pulse merges with the fast jet in Figures~6, 8, and~10.

The flow in the fast jet creates a terminal Mach disk shock near the
jet tip which reduces the jet gas velocity down to the flow velocity
of the contact discontinuity at the leading edge of the jet.  With
radiative cooling, the jet has a much higher density contrast near its
tip (as the shocked, heated gas cools radiatively, it compresses), a
much narrower bow shock, and lower average temperatures.

In Figures~4 and~5 at 21.1 yr (circa 2011) after the launch of the
initial bubble pulse, the jet has caught up with and penetrated almost
all the way across the nearly spherical bow shock bubble of the slower
initial pulse.  At times later than about 22 yr (see Figures~6 and~7),
the jet has penetrated through the bubble and thereafter begins to
subsume its spherical form.  The jet has absorbed the spherical bubble
almost entirely at 26.1 yr (circa 2016) in Figures~8 and~9, and
entirely at 30 yr (circa 2020) in Figures~10 and~11.  The jet is
clearly visible in [Fe~II] in Figures~5, 7, 9, and~11 with a strong
terminal shock at the tip of the jet and shocked gas knots throughout
the jet stem, while the cooler bow shock bubble is clearly visible in
H$_2$ in Figures~5 and~7, with the jet bow shock envelope visible
inside the bow shock bubble.  We believe Figures~6 and~7 (circa 2012)
coincide most closely with the 2011 Keck image in Figure~1 and the
related images in Hodapp \& Chini (2014).  Note that we have assumed
that the bow shock bubble is transparent to the [Fe~II] emission from
the jet; in actuality there may be some obscuring of the [Fe~II]
emission by the surrounding bow shock bubble, as suggested by
Figure~1.

There are as yet no detailed observational surface brightness maps for
the SVS 13 jet and bubble, but the data presented in Figure~1 indicate
that the maximum surface brightness of the jet is $S_{max} \sim 5
\times 10^{-13}$ erg cm$^{-2}$ arcsec$^{-2}$ s$^{-1}$ and that of the
bubble is $S_{max} \sim 10^{-13}$ erg cm$^{-2}$ arcsec$^{-2}$
s$^{-1}$.  These estimates are $5 \times$ and $10 \times$ brighter,
respectively, than the simulated maximum surface brightnesses for the
jet and bubble.  However it is not atypical for simulated surface
brightnesses to be an order of magnitude or so lower than observed
surface brightnesses (see for example Figure~8 of Tesileanu et
al.\ 2014).

There is one major difference between the simulated surface brightness
maps and Figure~1: the brightest part of the simulated H$_2$ emission
maps is the bow shock of the jet, rather than the bow shock bubble.
It is difficult though to disentangle the observational signals from
the jet bow shock and the bow shock bubble in Figure~1.  Furthermore
our model may simply produce a bow shock bubble that is too dim, and
additional physics might brighten it.

Post 2016 the SVS 13 jet should propagate as a standard proto-stellar
jet as in Figures~8 and~10 until the jet source turns off, and another
cycle is later begun.

The jet appears to be surrounded in H$_2$ emission in the Keck images,
involving entrainment of slower ambient gas by turbulent interaction
with the jet.  The broader envelope of the jet in the simulations
seems to be a reasonable representation of this turbulent interface.

\section{Conclusion}

We performed parallelized simulations using the WENO method of the
SVS 13 micro-jet and bow shock bubble which reproduce the main
features and dynamics of the Keck/OSIRIS data (Hodapp \& Chini 2014).

The simulated bow shock bubble has a diameter of approximately
$10^{10}$ km, which agrees with the Keck images.  For the simulations,
we found that good agreement with the Keck data was obtained if the
pulse and jet are half as dense as the near-field ambient, and twice
as hot.  Temperatures are kept well below 10,000~K for the most part
by assuming the bubble and jet are less dense than the immediate
ambient gas.  The center of the bow shock bubble propagates at
approximately 50 km~s$^{-1}$ over the first 23 yr of the simulations,
as observed in the Keck data.

The Keck images and data in Hodapp and Chini (2014) present an
``inverse'' problem: the Keck observations indicate an expanding
cooler bow shock bubble, with the bubble center moving at
approximately 50 km~s$^{-1}$ with a radial expansion velocity of 11
km~s$^{-1}$, surrounding the fast hotter jet, which is propagating at
156 km~s$^{-1}$.  In this paper, we reproduced those observations
through numerical simulations of an initial short jet pulse in 1990
which produces the spherical bow shock bubble, followed by a fast
continuous jet launched in 2008.  The slow cool spherical bubble plus
fast continuous jet cannot be modeled by a single jet---the single jet
would always look similar to Figures~8 and~10, except without any
trace of the spherical bubble.  The simulations of the initial bubble
pulse and later continuous jet provide good agreement with the Keck
data when the pulse and jet are about half as dense as the near-field
ambient (which derives from previous outflows), and twice as hot.

In the simulations, contact and bow shock waves are visible from both
the initial short jet pulse which creates the nearly spherical bow
shock bubble and from the fast jet; a terminal Mach disk shock is
visible near the tip of the continuous jet, which reduces the jet gas
velocity down to the flow velocity of the contact discontinuity at the
leading edge of the jet. The jet has caught up with and penetrated
almost all the way across the nearly spherical bow shock bubble of the
slower initial pulse at 21.1 yr (circa 2011).  At times later than
about 22 yr, the jet has penetrated through the bubble and thereafter
begins to subsume its spherical form.  The jet has absorbed the
spherical bubble almost entirely at 26.1 yr (circa 2016).  We believe
Figures~6 and~7 (circa 2012) correspond most closely with the 2011
Keck image in Figure~1 and the related images in Hodapp \& Chini
(2014).

The Keck image shows an asymmetric bow shock bubble, with a much
weaker shock on one side.  This weakening could be caused by the jet
running into heavier and cooler gas (see Figure~12) from a previous
outflow on that side.\\ \ \\

\noindent
{\bf Acknowledgment.} We would like to thank Perry Vargas for
extracting the line emissivity data from {\em Cloudy}.

\newpage

\section*{References}

\begin{itemize}

\item[] 
Draine, B.~T. 1980, {\em ApJ}, 241, 1021

\item[] 
Ferland, G.~J., Porter, R.~L., van Hoof, P.~A.~M., et al.\ 2013,
{\em RMxAA}, 49, 137

\item[] 
Gardner, C.~L., \& Dwyer, S.~J. 2009, {\em AcMaS}, 29B, 1677

\item[]
Ha, Y., Gardner, C.~L., Gelb, A., \& Shu, C.-W. 2005, {\em JSCom}, 24, 29

\item[]
Hodapp, K.~W., \& Chini, R. 2014, {\em ApJ}, 794, 169

\item[] 
Krist, J.~E., Stapelfeldt, K.~R., Hester, J.~J., et al.\ 2008, 
{\em AJ}, 136, 1980

\item[]
Le Bourlot, J., Pineau des For\^ets, G., \& Flower, D.~R. 1999,
{\em MNRAS}, 305, 802

\item[] 
Loinard, L., Zapata, L.~A., Rodr\'iguez, L. F., et al. 2013, 
{\em MNRAS}, 430, L10

\item[]
Reipurth, B., \& Bally, J. 2001, {\em ARA\&A}, 39, 403

\item[]
Schmutzler, T., \& Tscharnuter, W.~M. 1993, {\em A\&A}, 273, 318

\item[]
Shu, C.-W. 1999,
in {\em High-Order Methods for Computational Physics}, Lecture Notes
in Computational Science and Engineering vol.~9, (New York: Springer
Verlag), 439--582

\item[] 
Tesileanu, O., Matsakos, T., Massaglia, S., et al.\ 2014, {\em A\&A}, 562, A117

\end{itemize}

\newpage

\begin{figure}[htbp]
\begin{center}
\vspace{-3in}
\hspace*{-0.75in}\includegraphics[scale=0.8]{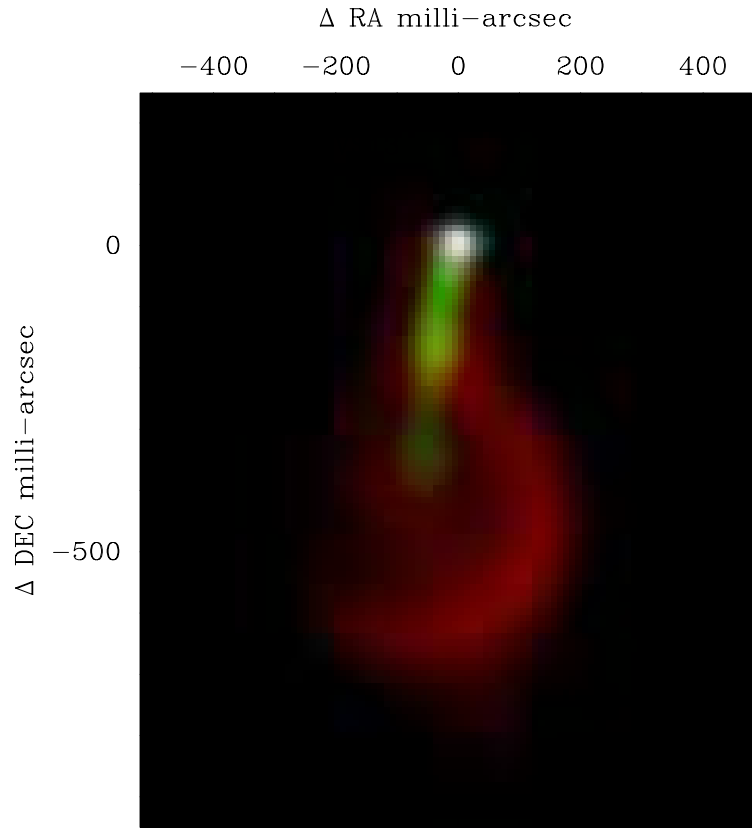}
\vspace{-1in}
\caption{\label{fig-1} Keck Telescope laser adaptive-optics integral
  field spectroscopy image with OSIRIS by K.~W. Hodapp of SVS 13 on
  2011 August 21.  The micro-jet is traced by the emission of the
  shock excited 1.644 $\mu$m [Fe~II] line (green) and the bow shock
  bubble is traced in the lower excitation 2.122 $\mu$m H$_2$ 1--0
  S(1) line (red).  The white disk is the stellar source.  The
  micro-jet plus bubble region is approximately $10^{10}$ km on each
  side.}
\end{center}
\end{figure}

\begin{figure}[htbp]
\begin{center}
\includegraphics[scale=0.6]{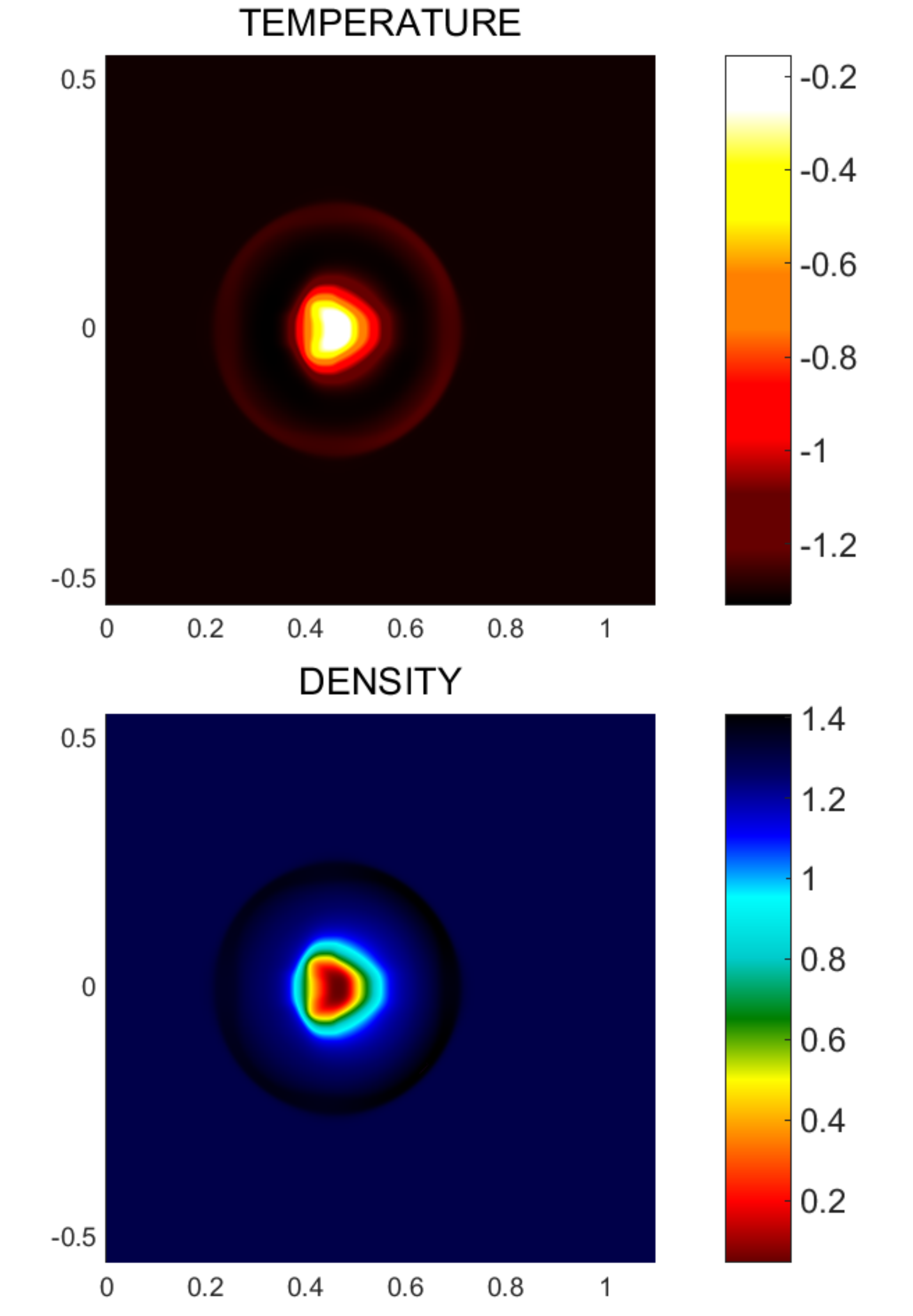}
\caption{\label{fig-T-rho1} Cross section of the numerical simulation
  of the SVS 13 bow shock bubble at 18.5 yr (circa 2008), just before
  the fast jet is turned on: logarithm of temperature
  $\log_{10}\left(T/\overline{T}\right)$, where $\overline{T}$ =
  $10^4$ K, and logarithm of density
  $\log_{10}\left(\rho/\overline{\rho}\right)$, where
  $\overline{\rho}$ = 100 H cm$^{-3}$.  Lengths along the left and
  bottom boundaries are in $10^{10}$ km for Figures~2--13.  Note that
  all simulation images are more or less zoomed in.}
\end{center}
\end{figure}

\begin{figure}[htbp]
\begin{center}
\includegraphics[scale=0.6]{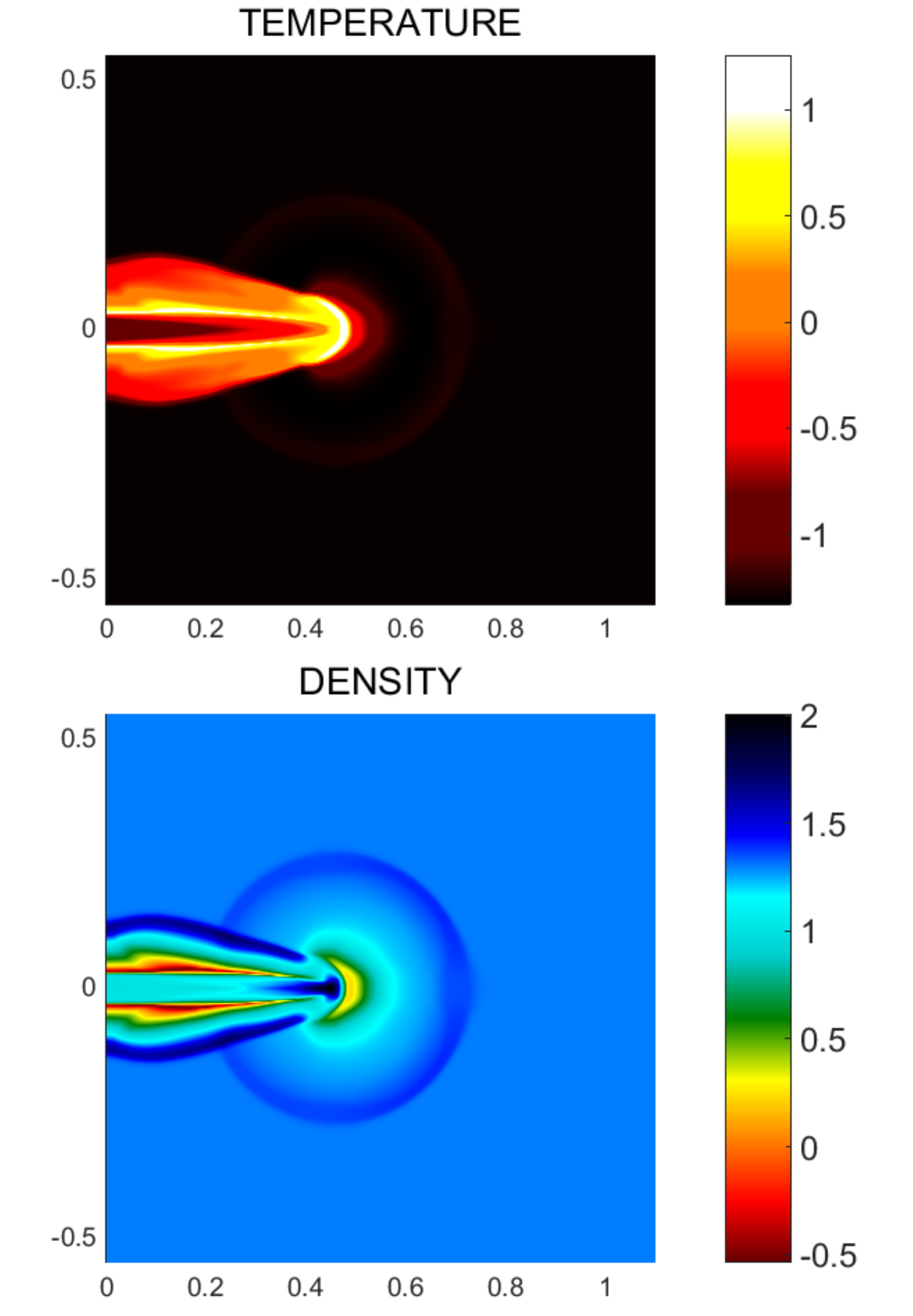}
\caption{\label{fig-T-rho2} Cross section of the numerical simulation
  of the SVS 13 bow shock bubble and jet at 20.4 yr (circa 2010):
  logarithm of temperature $\log_{10}\left(T/\overline{T}\right)$,
  where $\overline{T}$ = $10^4$ K, and logarithm of density
  $\log_{10}\left(\rho/\overline{\rho}\right)$, where
  $\overline{\rho}$ = 100 H cm$^{-3}$.  Lengths are in $10^{10}$ km.}
\end{center}
\end{figure}

\begin{figure}[htbp]
\begin{center}
\includegraphics[scale=0.6]{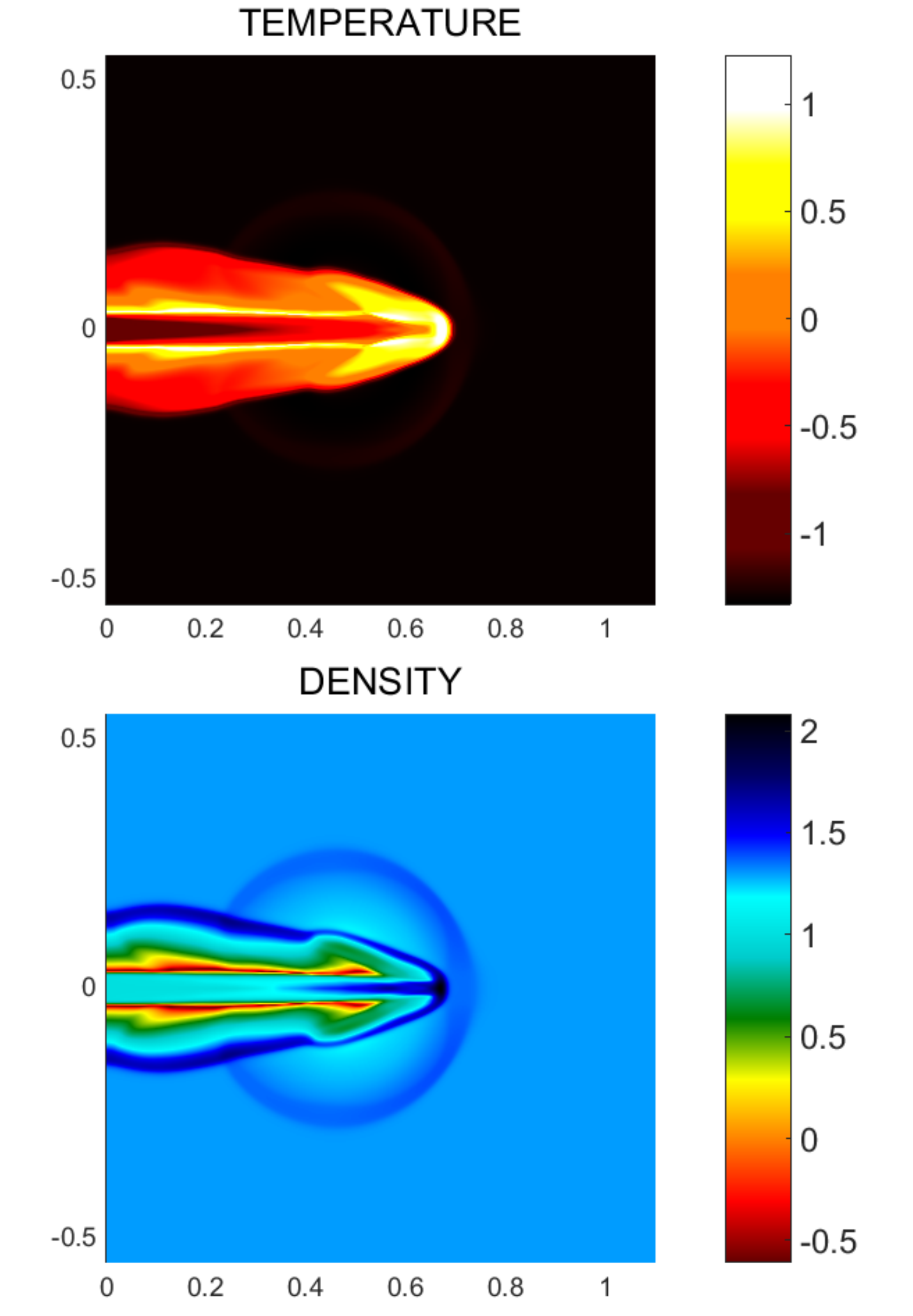}
\caption{\label{fig-T-rho3} Cross section of the numerical simulation
  of the SVS 13 bow shock bubble and jet at 21.1 yr (circa 2011):
  logarithm of temperature $\log_{10}\left(T/\overline{T}\right)$,
  where $\overline{T}$ = $10^4$ K, and logarithm of density
  $\log_{10}\left(\rho/\overline{\rho}\right)$, where
  $\overline{\rho}$ = 100 H cm$^{-3}$.  Lengths are in $10^{10}$ km.}
\end{center}
\end{figure}

\begin{figure}[htbp]
\begin{center}
\includegraphics[scale=0.6]{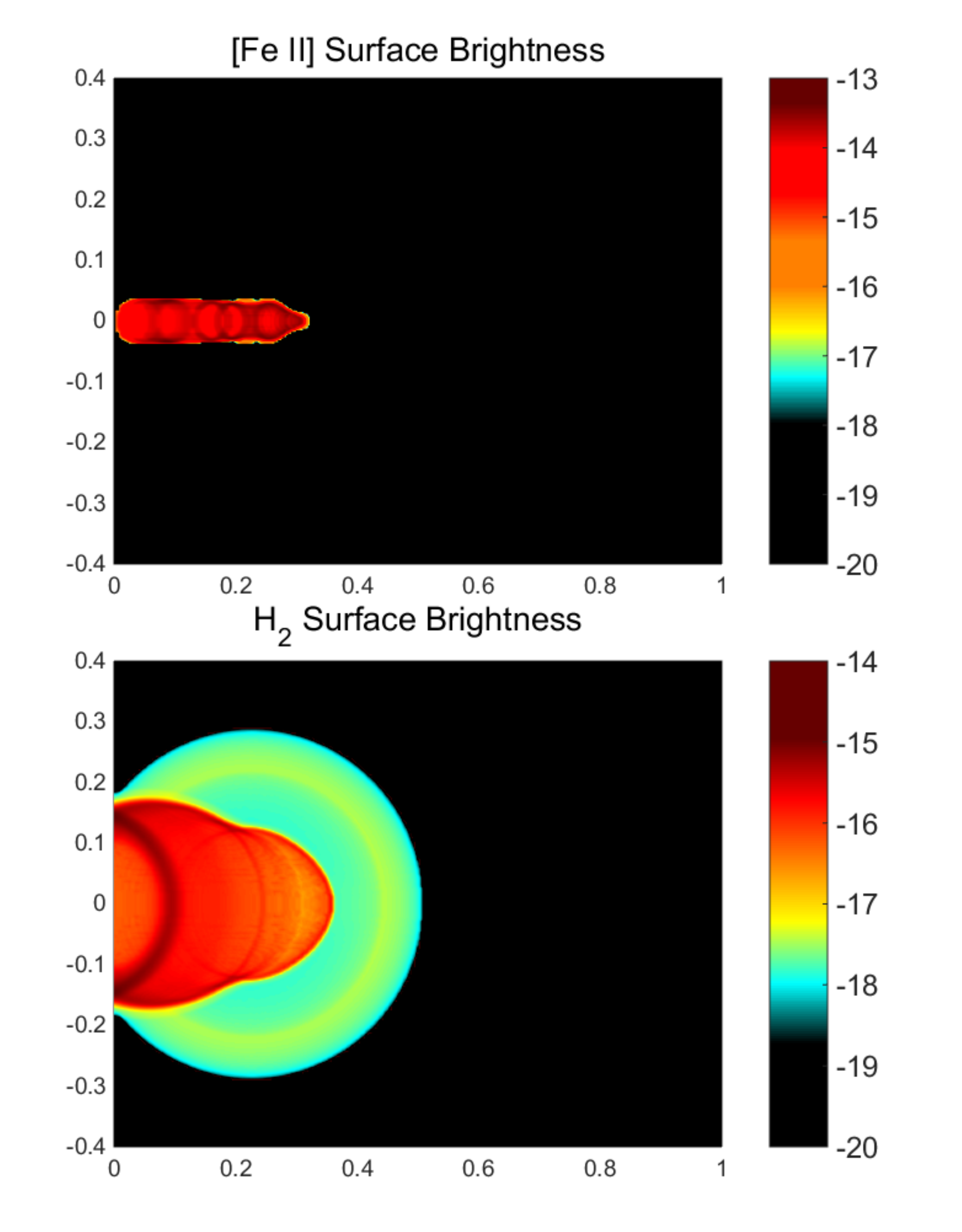}
\caption{\label{fig-lines2} Simulated surface brightness $S$ in the
  1.644 $\mu$m [Fe~II] line and the 2.122 $\mu$m H$_2$ line of the SVS
  13 jet and bow shock bubble at 21.1 yr (circa 2011): $\log_{10}(S)$
  with $S$ in erg cm$^{-2}$ arcsec$^{-2}$ s$^{-1}$.  The jet and
  bubble have been projected with a declination angle of 30$^\circ$
  with respect to the line of sight.  Lengths are in $10^{10}$ km.}
\end{center}
\end{figure}

\begin{figure}[htbp]
\begin{center}
\includegraphics[scale=0.6]{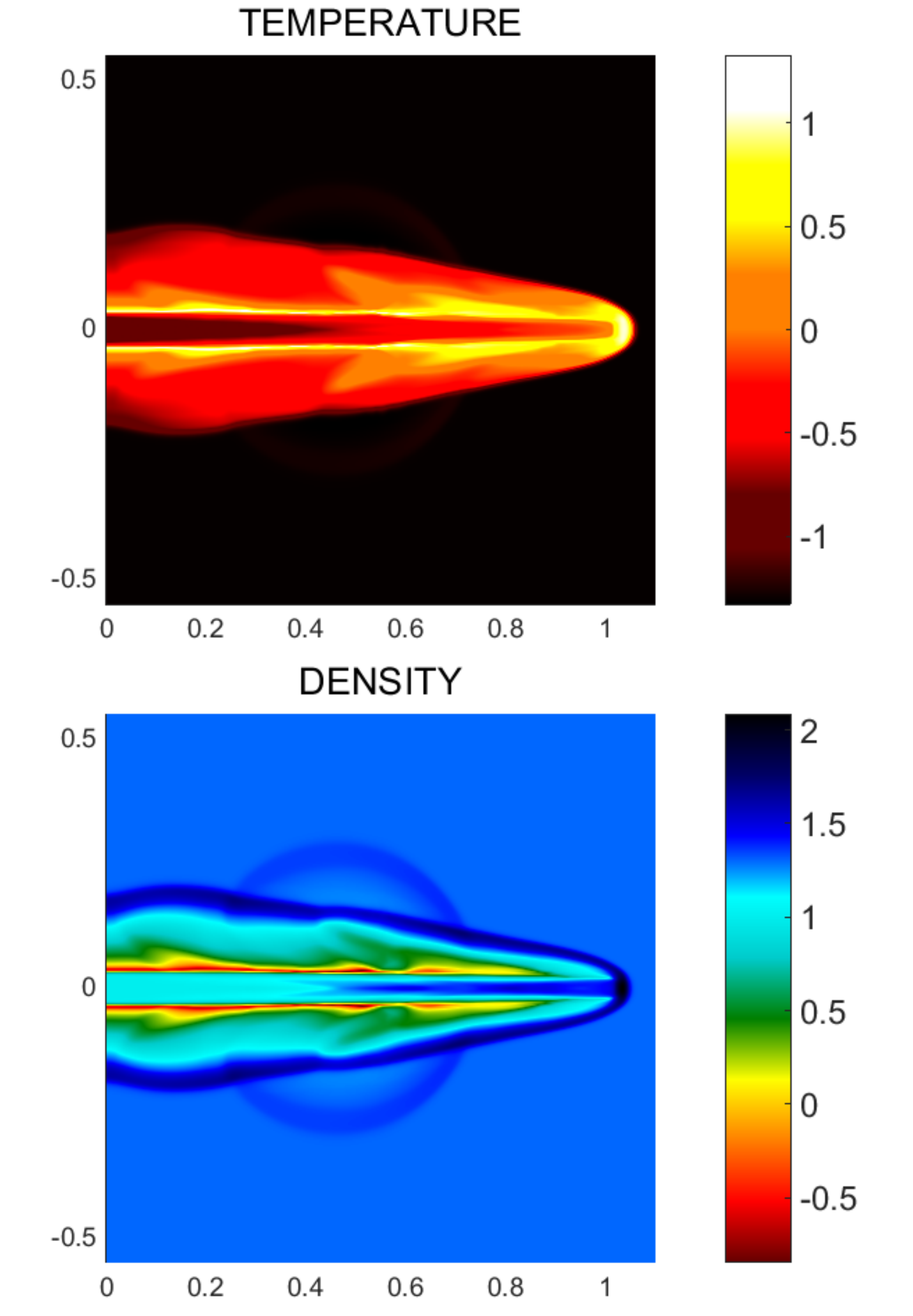}
\caption{\label{fig-T-rho4} Cross section of the numerical simulation
  of the SVS 13 bow shock bubble and jet at 22.3 yr (circa 2012):
  logarithm of temperature $\log_{10}\left(T/\overline{T}\right)$,
  where $\overline{T}$ = $10^4$ K, and logarithm of density
  $\log_{10}\left(\rho/\overline{\rho}\right)$, where
  $\overline{\rho}$ = 100 H cm$^{-3}$.  Lengths are in $10^{10}$ km.}
\end{center}
\end{figure}

\begin{figure}[htbp]
\begin{center}
\includegraphics[scale=0.6]{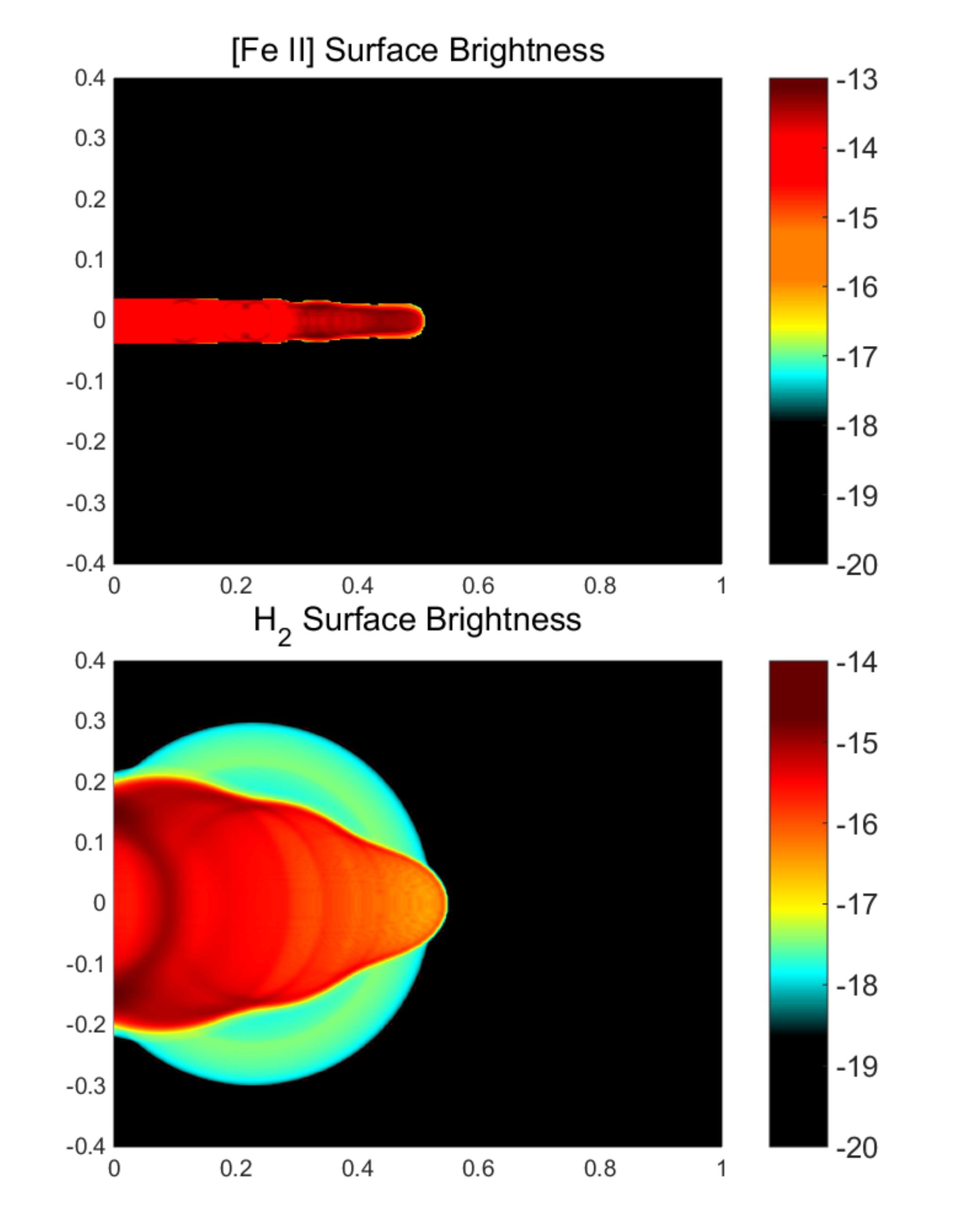}
\caption{\label{fig-lines3} Simulated surface brightness $S$ in the
  1.644 $\mu$m [Fe~II] line and the 2.122 $\mu$m H$_2$ line of the SVS
  13 jet and bow shock bubble at 22.3 yr (circa 2012): $\log_{10}(S)$
  with $S$ in erg cm$^{-2}$ arcsec$^{-2}$ s$^{-1}$.  The jet and
  bubble have been projected with a declination angle of 30$^\circ$
  with respect to the line of sight.  Lengths are in $10^{10}$ km.}
\end{center}
\end{figure}

\begin{figure}[htbp]
\begin{center}
\includegraphics[scale=0.6]{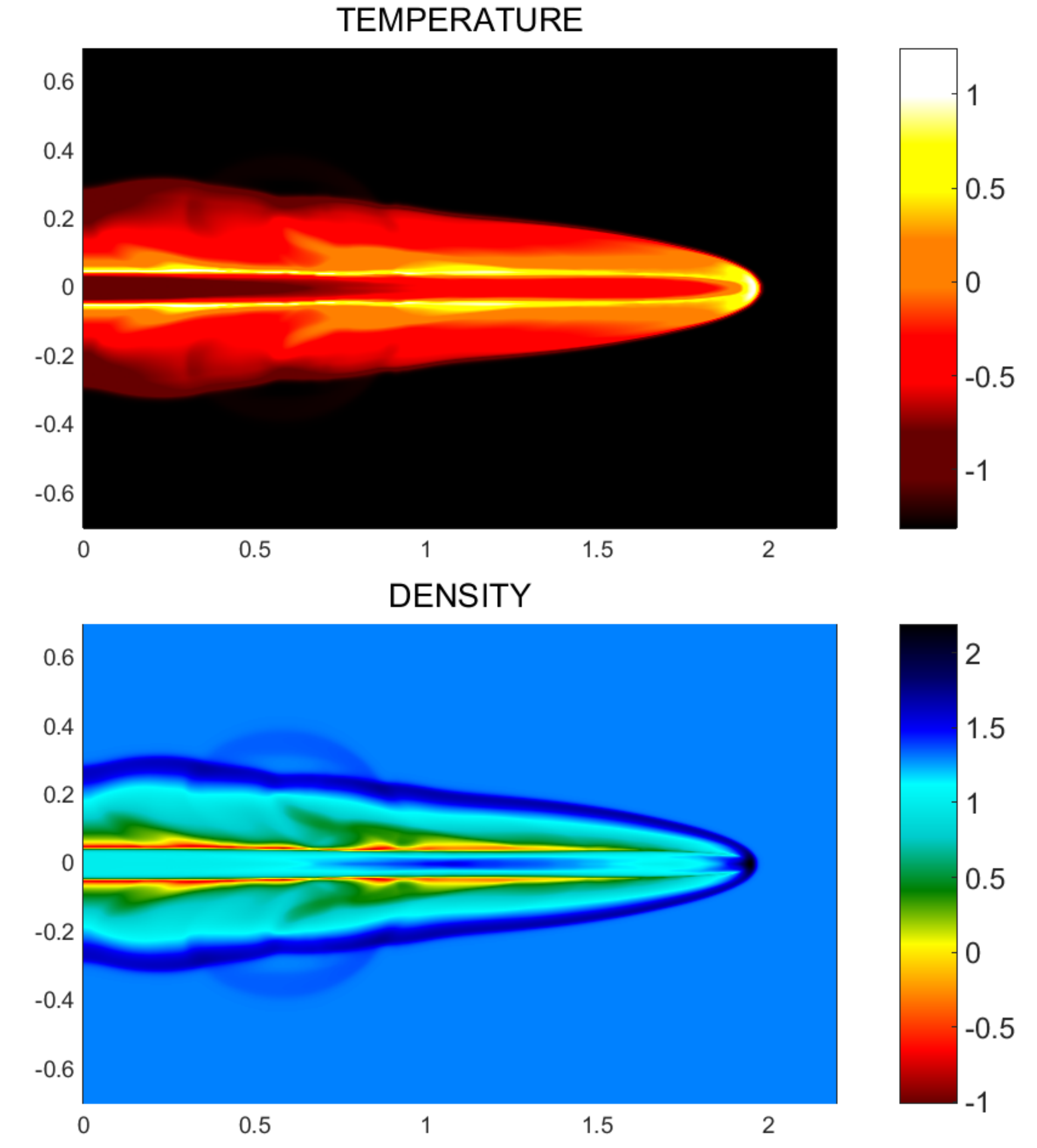}
\caption{\label{fig-T-rho5} Cross section of the numerical simulation
  of the SVS 13 bow shock bubble and jet at 26.1 yr (circa 2016),
  assuming the jet source has not turned off: logarithm of temperature
  $\log_{10}\left(T/\overline{T}\right)$, where $\overline{T}$ =
  $10^4$ K, and logarithm of density
  $\log_{10}\left(\rho/\overline{\rho}\right)$, where
  $\overline{\rho}$ = 100 H cm$^{-3}$.  Lengths are in $10^{10}$ km.}
\end{center}
\end{figure}

\begin{figure}[htbp]
\begin{center}
\includegraphics[scale=0.6]{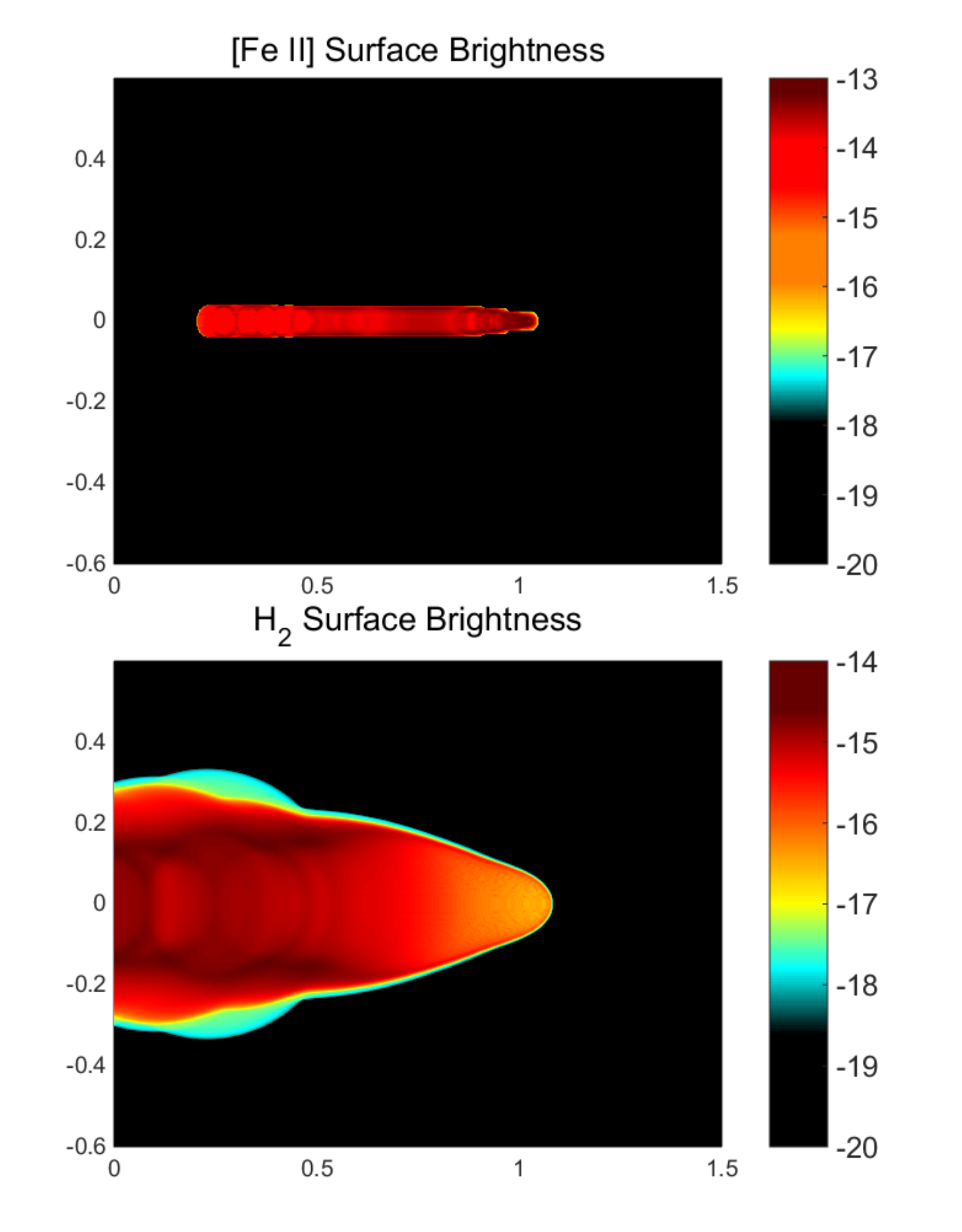}
\caption{\label{fig-lines4} Simulated surface brightness $S$ in the
  1.644 $\mu$m [Fe~II] line and the 2.122 $\mu$m H$_2$ line of the SVS
  13 jet and bow shock bubble at 26.1 yr (circa 2016): $\log_{10}(S)$
  with $S$ in erg cm$^{-2}$ arcsec$^{-2}$ s$^{-1}$.  The jet and
  bubble have been projected with a declination angle of 30$^\circ$
  with respect to the line of sight.  Lengths are in $10^{10}$ km.}
\end{center}
\end{figure}

\begin{figure}[htbp]
\begin{center}
\includegraphics[scale=0.6]{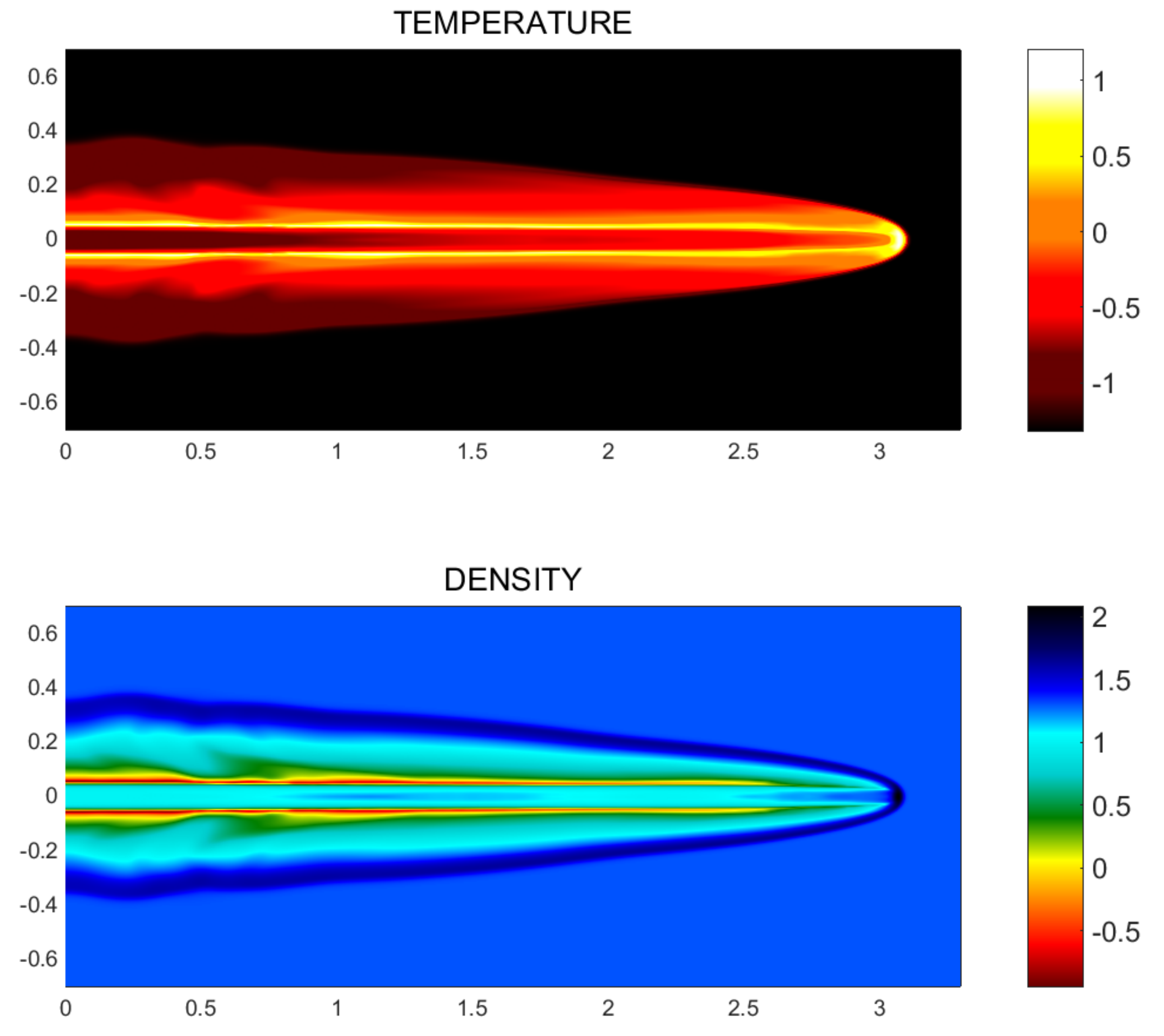}
\caption{\label{fig-T-rho6} Cross section of the numerical simulation
  of the SVS 13 bow shock bubble and jet at 30 yr (circa 2020),
  assuming the jet source has not turned off: logarithm of temperature
  $\log_{10}\left(T/\overline{T}\right)$, where $\overline{T}$ =
  $10^4$ K, and logarithm of density
  $\log_{10}\left(\rho/\overline{\rho}\right)$, where
  $\overline{\rho}$ = 100 H cm$^{-3}$.  Lengths are in $10^{10}$ km.}
\end{center}
\end{figure}

\begin{figure}[htbp]
\begin{center}
\includegraphics[scale=0.6]{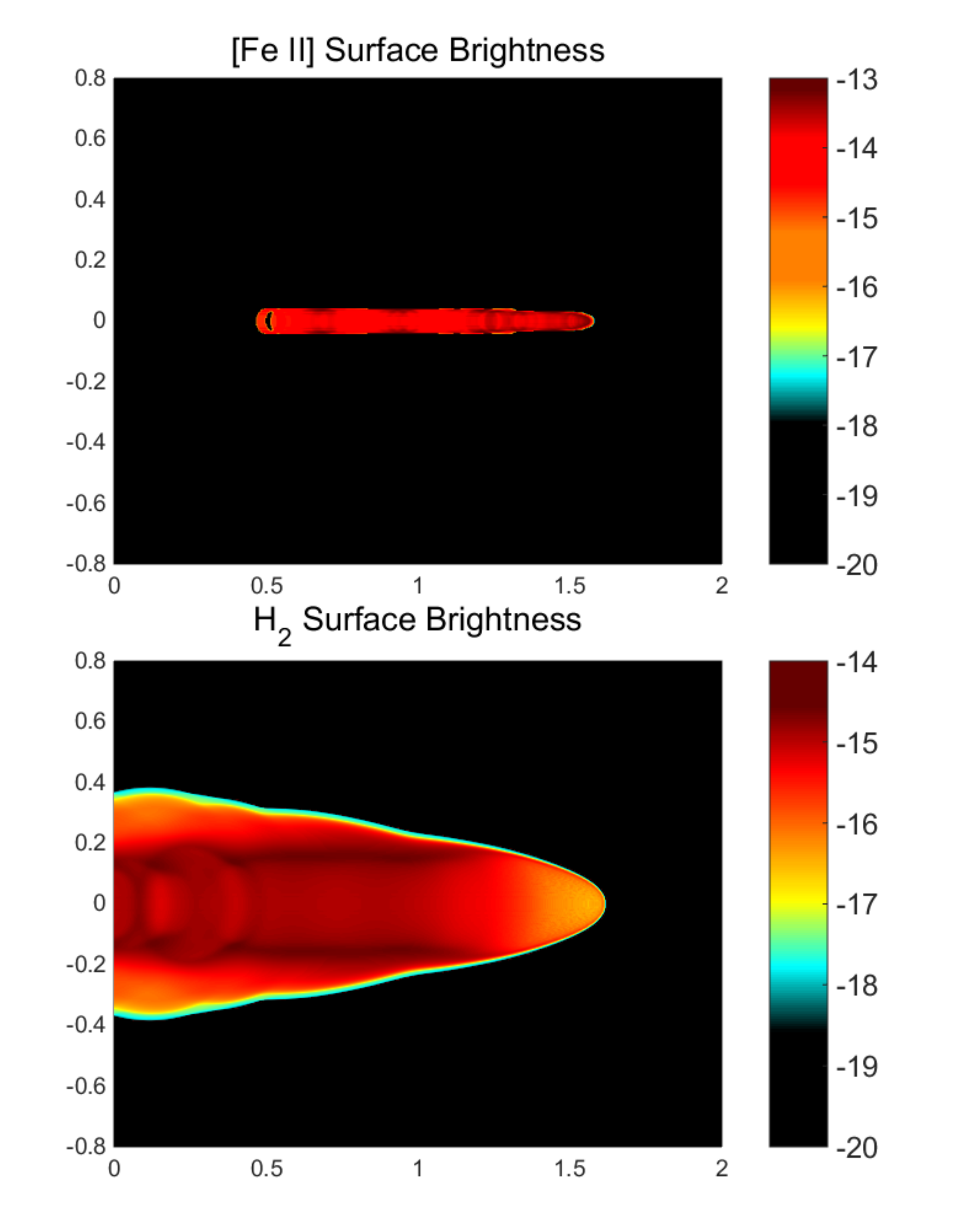}
\caption{\label{fig-lines5} Simulated surface brightness $S$ in the
  1.644 $\mu$m [Fe~II] line and the 2.122 $\mu$m H$_2$ line of the SVS
  13 jet and bow shock bubble at 30 yr (circa 2020): $\log_{10}(S)$
  with $S$ in erg cm$^{-2}$ arcsec$^{-2}$ s$^{-1}$.  The jet and
  bubble have been projected with a declination angle of 30$^\circ$
  with respect to the line of sight.  Lengths are in $10^{10}$ km.}
\end{center}
\end{figure}

\begin{figure}[htbp]
\begin{center}
\includegraphics[scale=0.6]{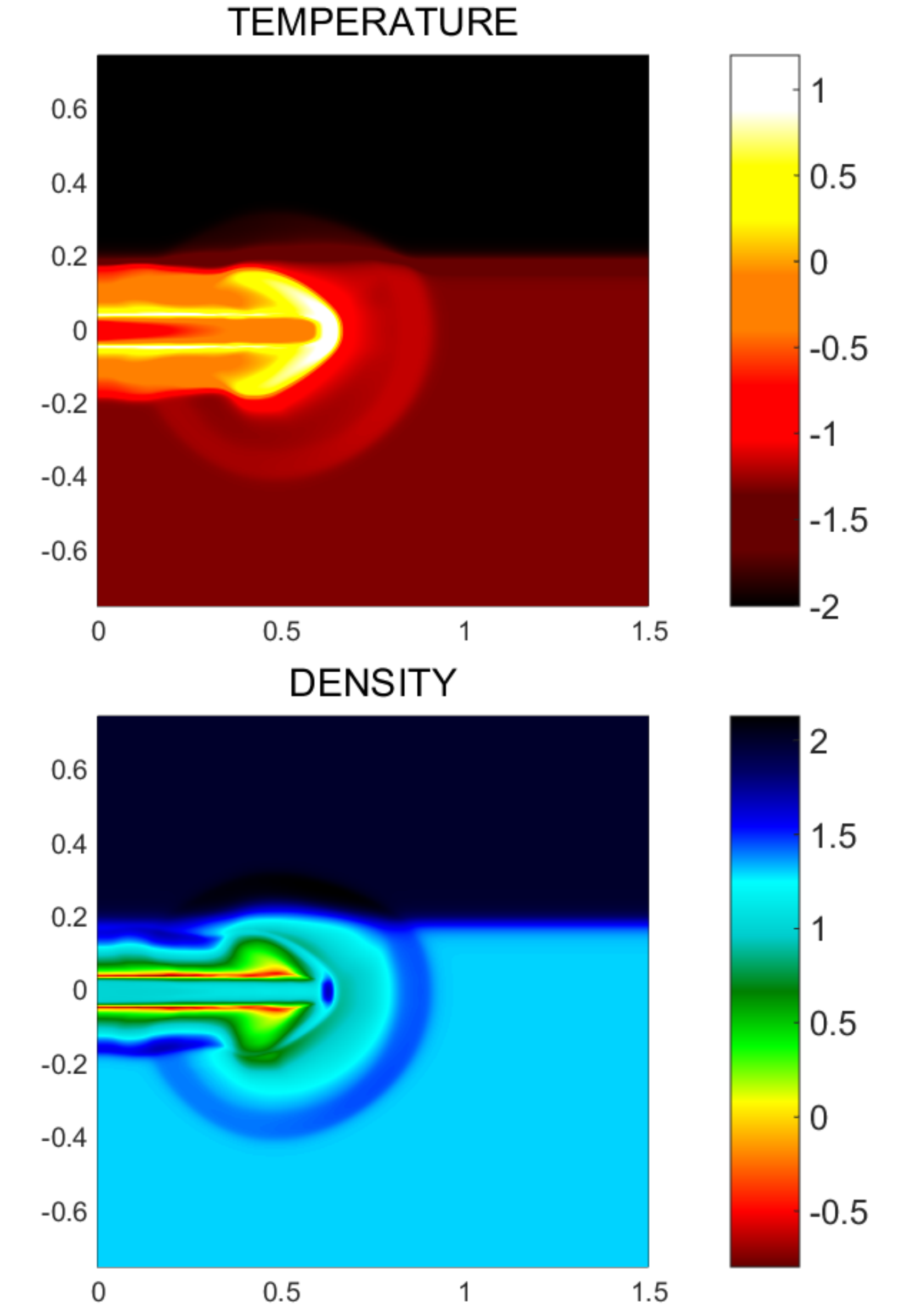}
\caption{\label{fig-T-rho-heavy} Cross section of the numerical
  simulation of the SVS 13 bow shock bubble and jet at 21 yr with a
  heavier ambient (5 times the usual $\rho_a$) ``above'' the jet:
  logarithm of temperature $\log_{10}\left(T/\overline{T}\right)$,
  where $\overline{T}$ = $10^4$ K, and logarithm of density
  $\log_{10}\left(\rho/\overline{\rho}\right)$, where
  $\overline{\rho}$ = 100 H cm$^{-3}$.  Lengths are in $10^{10}$ km.}
\end{center}
\end{figure}

\begin{figure}[htbp]
\begin{center}
\includegraphics[scale=0.6]{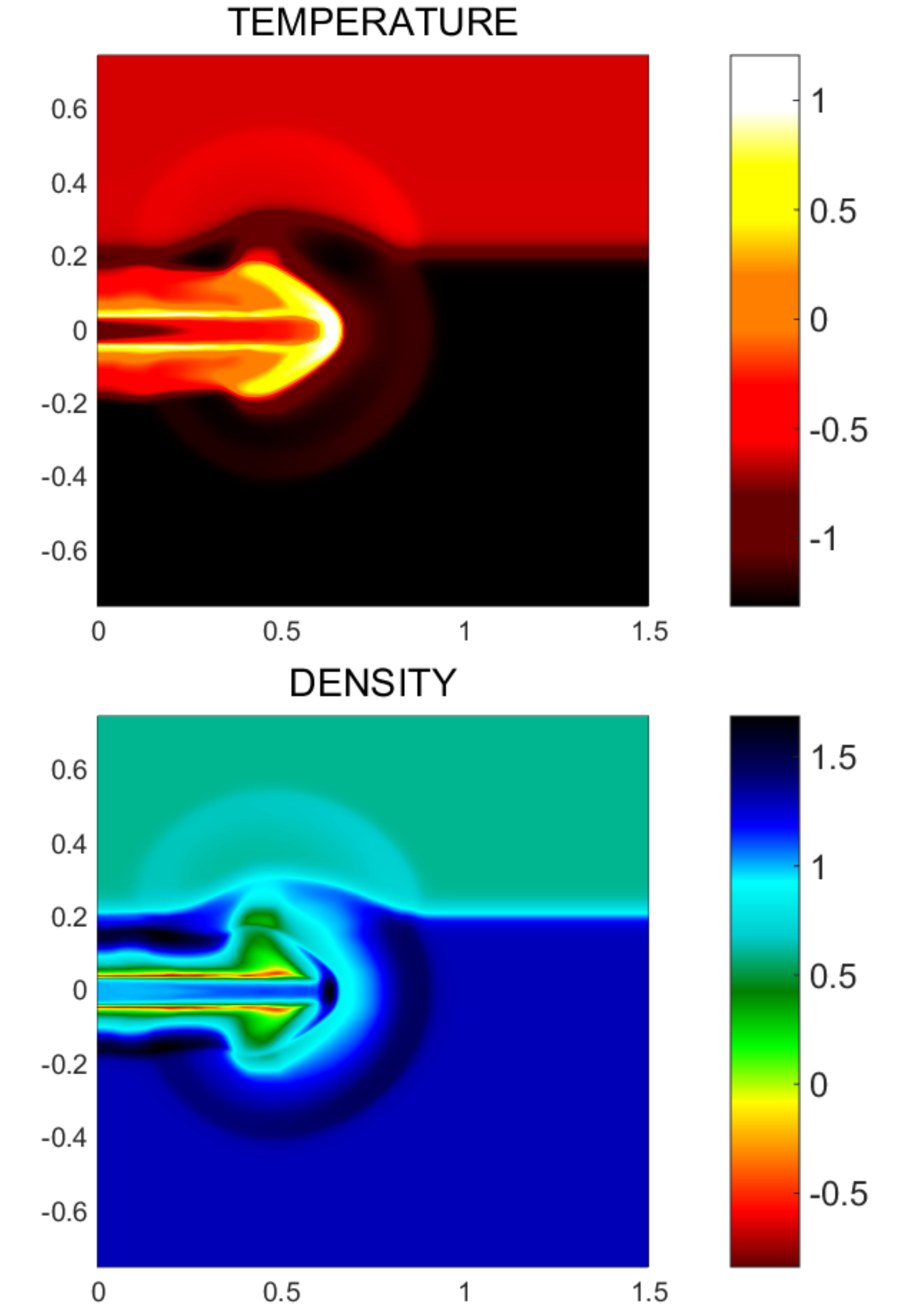}
\caption{\label{fig-T-rho-light} Cross section of the numerical
  simulation of the SVS 13 bow shock bubble and jet at 21 yr with a
  lighter ambient (1/5 the usual $\rho_a$) ``above'' the jet:
  logarithm of temperature $\log_{10}\left(T/\overline{T}\right)$,
  where $\overline{T}$ = $10^4$ K, and logarithm of density
  $\log_{10}\left(\rho/\overline{\rho}\right)$, where
  $\overline{\rho}$ = 100 H cm$^{-3}$.  Lengths are in $10^{10}$ km.}
\end{center}
\end{figure}

\end{document}